\documentstyle[epsfig]{mn}
\begin{document}
\def\ltsima{$\; \buildrel < \over \sim \;$}
\def\simlt{\lower.5ex\hbox{\ltsima}}
\def\gtsima{$\; \buildrel > \over \sim \;$}
\def\simgt{\lower.5ex\hbox{\gtsima}}
\def\approxgt{\mathrel{\hbox{\rlap{\lower.55ex \hbox {$\sim$}}
        \kern-.3em \raise.4ex \hbox{$>$}}}}
\def\approxlt{\mathrel{\hbox{\rlap{\lower.55ex \hbox {$\sim$}}
        \kern-.3em \raise.4ex \hbox{$<$}}}}

\title[]{Discovery of two new Fast X-ray Transients with INTEGRAL: IGR~J03346+4414 and IGR~J20344+3913} 

\author[Sguera et al.]
{V. Sguera$^{1}$, L. Sidoli$^2$, A. Paizis$^2$, A. J. Bird$^3$
\\ 
$^1$ INAF, Istituto di Astrofisica Spaziale e Fisica Cosmica, Via Gobetti 101, I-40129 Bologna, Italy \\
$^2$ INAF, Istituto di Astrofisica Spaziale e Fisica Cosmica, Via E. Bassini  15, I-20133 Milano, Italy \\
$^3$ School of Physics and Astronomy, University of Southampton, University Road, Southampton, SO17 1BJ, UK \\
}

\date{Accepted 2016 August 26.  In original form 2016 June 1}

\maketitle 

\begin{abstract}
We report on the discovery of two Fast X-ray Transients (FXTs) from analysis of archival INTEGRAL data. Both  are characterized by a remarkable hard X-ray activity above 20 keV,  in term of duration ($\sim$ 15 and 30 minutes, respectively), peak-flux ($\sim$ 10$^{-9}$  erg cm$^{-2}$ s$^{-1}$) and  dynamic range ($\sim$ 2400 and 1360,  respectively). Swift/XRT follow-up observations  failed to detect any quiescent or low level  soft X-ray emission from either of  the two FXTs, providing  an upper limit  of the order of a few times 10$^{-12}$  erg cm$^{-2}$ s$^{-1}$.  The main spectral and temporal IBIS/ISGRI characteristics are presented and discussed  with the aim of infering possible hints on their 
nature. 
\end{abstract}


\vspace{1.0cm}

\section{Introduction}

Since its launch in 2002, the IBIS/ISGRI detector  (Lebrun et al. 2003, Ubertini et al. 2003) on board the INTEGRAL observatory (Winkler et al. 2003) 
has  repeatedly  proven its suitableness for the investigation of  the hard X-ray transient sky 
at energies above 20 keV.  Thanks to its powerful combination of  large field of view (29$^\circ$$\times$29$^\circ$, partially coded at zero response),  fine  angular resolution (12$'$) and  good instantaneous sensitivity ($\sim$ 7 mCrab at 2$\sigma$ level  with an exposure of $\sim$ 2000 s), it has discovered several hundred new  hard X-ray transients both on short (hours--days) and long (weeks--months) timescales (e.g. Bird et al. 2016). The great majority of these sources  is Galactic  in origin since  INTEGRAL  has spent  a considerable fraction of its observational time towards the Galactic plane.  In particular,  one of the major  outcomes has been  the discovery of many new fast X-ray transients peculiarly characterized by activity on very short timescale (typically only a  few hours as detected above 20 keV) and subsequently   optically identified, decidedly unexpectedly,  as supergiant HMXBs  (SGXBs). This led to the recognition of a new class of SGXBs which has been named  Supergiant Fast X-ray Transients (SFXTs, Negueruela et al. 2006, Sguera et al. 2005, 2006). In fact, before the  INTEGRAL  era,  the $`$classical$`$ SGXBs were mistakenly assumed to all 
be bright persistent X-ray emitters. 

Fast X-ray transients (FXTs) are among the most elusive X-ray sources in the sky,  they are very difficult to discover because their activity  is particularly short and occurs at unpredictable locations and times. X-ray instruments having a sufficiently wide field of view and good instantaneous   sensitivity, such as IBIS/ISGRI, have the best chance of serendipitously detecting such short duration random events.  Due to their elusive nature, a large population of still undetected FXTs  could be hidden in our Galaxy;  it seems plausible that many such sources wait to be discovered. 

Here we report on the  analysis of  archival INTEGRAL data pertaining to observations of  specific regions of the Galactic plane (i.e. Cygnus region and Galactic anticenter) with the aim of finding  new FXTs.  As result, we report  on the discovery of two new such sources which have not been previously  detected by any other X-ray telescope. Both  are characterized by short (15--30 minutes duration)  and bright (average hard X-ray flux of $\sim$ 10$^{-9}$  erg cm$^{-2}$ s$^{-1}$)  outbursts as detected by INTEGRAL. Their main spectral and temporal characteristics are presented and discussed with the aim of infering possible hints on their nature.

\section{{\itshape INTEGRAL}}

\subsection{Data analysis}

\begin{table}
\caption {Log of IBIS/ISGRI observations} 
\label{tab:main_outbursts} 
\begin{tabular}{lccc}
\hline
\hline   
telescope   orbit                              &  range time        &    ScWs        &    exp        \\
                                                    &                                &        (n$^{\circ}$)                     &              (Ms)        \\
\hline    
\textbf{Anticenter}                         &                              &                              &                         \\ 
\textbf{(l$\sim$ 150$^\circ$)}                     &                            &                               &                   \\ 
960--966                                         &    Aug-Sep 2010                        &  569                   &      $\sim$  1  \\   
1150                                             &       Mar 2012                       &           44                   &        $\sim$   0.1   \\   
1199--1207, 1212                         &      Aug-Sep 2012                          &  655                   &      $\sim$ 1.3     \\   
1254--1255                                  &       Jan 2013                            &  70                   &      $\sim$   0.15     \\    
1260--1262                                   &        Feb 2013                         &   146                  &      $\sim$  0.3   \\   
\textbf{(l$\sim$ 225$^\circ$)}                 &                                                   &                           &                            \\ 
918--924                                      &        Apr-May 2010                   &   590                  &      $\sim$    1.2     \\   
978--980, 982                              &       Oct 2010                        &   218                  &      $\sim$      0.45       \\     
\hline        
                                                    &                                              &   2,292                       &      $\sim$ 4.5   \\  
\hline
\hline   
 \textbf{Cygnus}                         &                                    &                              &                            \\ 
1535--1537, 1539                &              Apr-May 2015                           &        118                     &         $\sim$  0.25             \\   
1541--1543                        &       May 2015                                      &             185                &         $\sim$      0.4        \\   
1554--1563                        &         Jun-Jul 2015                                     &        464                     &       $\sim$    1.6             \\   
1600--1603                      &          Oct 2015                                   &                171             &           $\sim$     0.35        \\   
1605, 1607                     &               Nov 2015                                  &            113                 &         $\sim$  0.25              \\   
1609--1610, 1614              &           Nov 2015                                  &            124                 &          $\sim$ 0.25             \\   
1616, 1621, 1624          &                Dec 2015                                &              123               &          $\sim$   0.25    \\ 
1626--1629                      &               Dec 2015-Jan 2016                &              232               &           $\sim$  0.7            \\   
\hline
                                      &                                                                   &            1,530             &      $\sim$ 4   \\  
\hline
\hline  
\end{tabular}
\end{table}

For our study, we used data collected with the ISGRI detector which is  the lower energy 
layer of the IBIS coded mask telescope. The  reduction  and  analysis  of  the data  have been  performed  by  using  the  Offline  Scientific
Analysis  (OSA)  version 10.1. INTEGRAL  observations  in each telescope orbit ($''$revolution$''$) are  divided  into short pointings (Science Window, ScW) having  a typical duration of $\sim$ 2000 s. 

Our data set consists of  targeted observations of the Galactic anticenter region as well as of the  Cygnus region.  For the Galactic anticenter, two different  positions have been observed  at  l $\sim$ 150$^\circ$  and l $\sim$ 225$^\circ$ respectively,  for a total  exposure of $\sim$ 4.5 Ms (see details in Table 1). For the Cygnus region, the data set  amounts to a total exposure of $\sim$ 4 Ms (see Table 1 for details) and consists of  recent public ToO observations  of V404 Cyg  as well as recent public observations 
targeted on  the  Cygnus sky region.  
 
 We performed an analysis at the ScW level of the full data set 
to  search for new X-ray transient sources detected with a significance greater than at least 6$\sigma$ in a single 
ScW. The search was initially performed in the energy band 22--60 keV;  this choice takes into account the evolution 
of the IBIS/ISGRI energy threshold  that occurred from revolution  number  $\sim$ 900 on. When an interesting excess was found,  we also checked the detection at higher energies (i.e. 60--100 keV) or in other  different  ranges (i.e. 22--30, 30--60 and 22--40 keV). 
The  sensitivity limit for a persistent source detected at 5$\sigma$ level (22--60 keV) in only one ScW of
about 2000 s duration is  $\sim$ 18 mCrab (Krivonos et al. 2010). 
We  note that this approach is  efficient in  unveiling  FXTs with durations as short as a few tens of minutes, since the search occurs on the same timescale as their outburst activity  themselves. Integrating for longer periods just degrades
the signal-to-noise ratio of the detection below detectability. The resulting excesses
were carefully inspected to identify them as real sources. They were visually 
examined to ensure an appropriate point spread function and to reject false detections such as 
ghosts. In order to rule out an artificial nature due to background noise or  structures/artifacts, 
we inspected the overall ScW images rms and checked the residuals maps. 
The absence of any known or found systematic effect gives us confidence about
the  real  nature  of  the two newly discovered FXTs. 

A detailed timing and spectral analysis was performed for each newly discovered FXT.
Due to possible cross-talk between sources in the same Field of View (FoV), 
we have also investigated the variability pattern of all other
brighter sources in the FoV, besides  the two sources of interest. The latter
have shown a different time variability, enabling us to conclude
that their light curves are uncontaminated and reliable.

The X-ray monitor JEM--X   makes observations simultaneously with
IBIS/ISGRI, although with a much smaller FoV,  providing images in the softer energy band 3--35 keV. 
Unfortunately both the here reported  newly discovered FXTs  were outside the JEM--X FoV. 

Through the paper, the spectral analysis was performed using
XSPEC version 12.9.0 and, unless stated otherwise, errors are quoted at the 90 per cent confidence
level for one single parameter of interest.

\subsection{Results}

\subsubsection{IGR J03346+4414}

\begin{figure}
\begin{center}
\epsfig{file=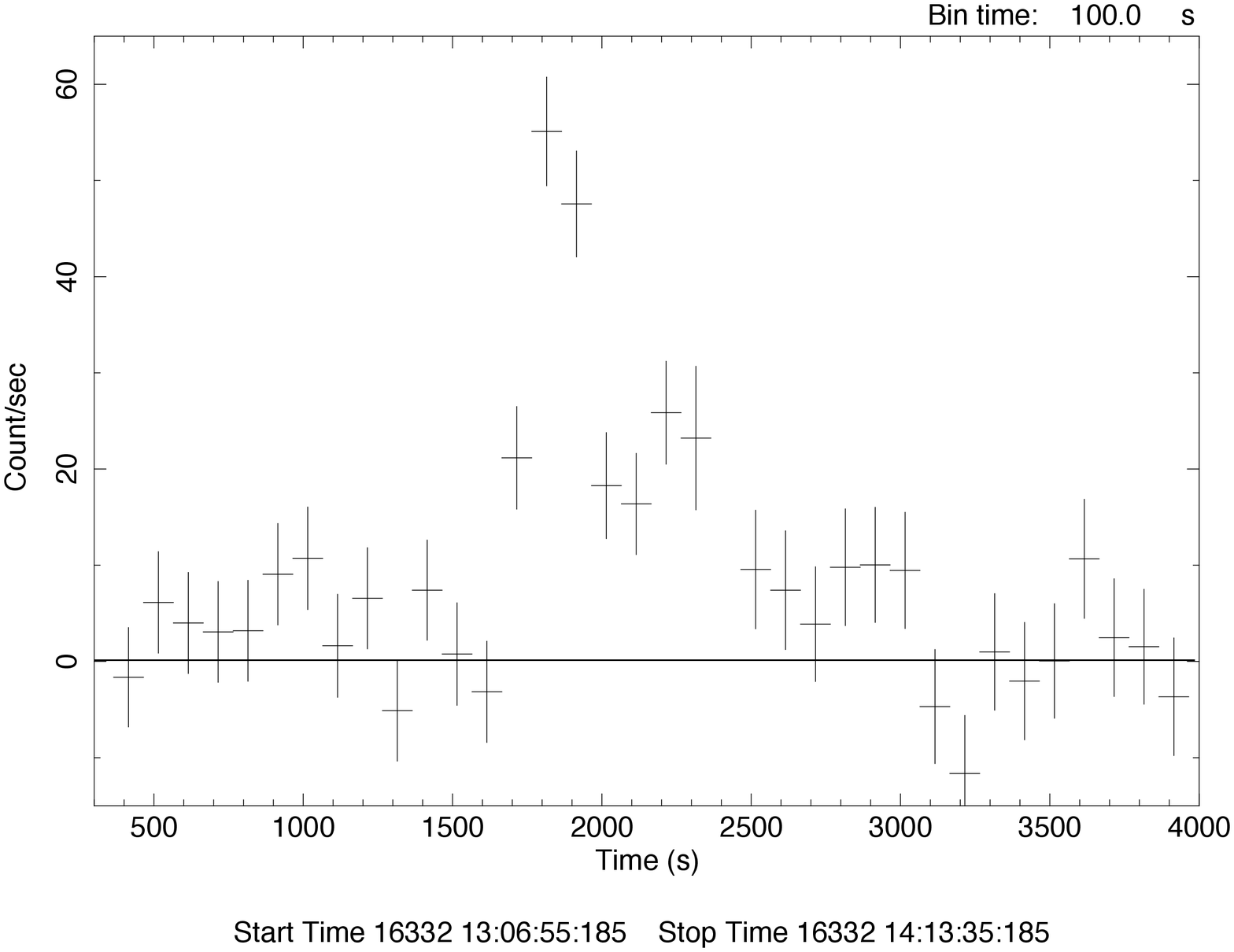,height=6.5cm,width=8.5cm}
\caption{IBIS/ISGRI 22--60 keV light curve (100 s bin time) of the newly discovered source IGR J03346+4414. For display purposes aimed at highlighting the transient nature of the source, the light curve was extracted from the two consecutive ScW number 10 and 11 (revolution 1261) although a source detection was obtained only from ScW number 10.}
\epsfig{file=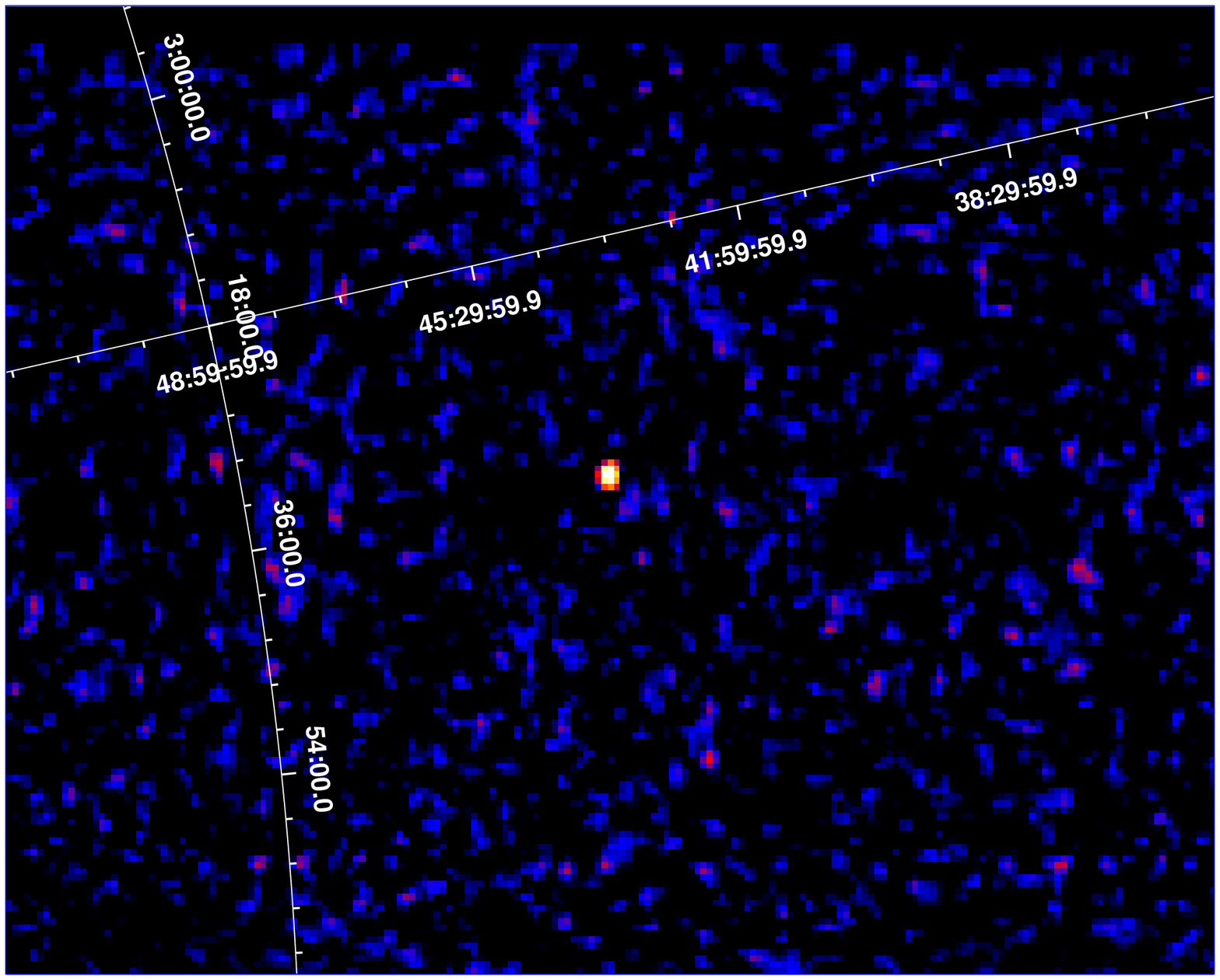,height=5.5cm,width=8.5cm}
 \caption{IBIS/ISGRI 22--60 keV significance image (ScW number 10, revolution 1261) of  the newly discovered source IGR J03346+4414. It was  detected with a significance of $\sim$ 12$\sigma$. }
\label{fig2}
\epsfig{file=fig3.ps,height=8.5cm,width=5.5cm, angle=-90}
 \caption{IBIS/ISGRI spectrum    of IGR J03346+4414 (extracted with the GTI from ScW number 10) fitted by a power law. The lower panel shows the residuals from the fit.}
\end{center}
\end{figure}

\begin{figure}
\begin{center}
\epsfig{file=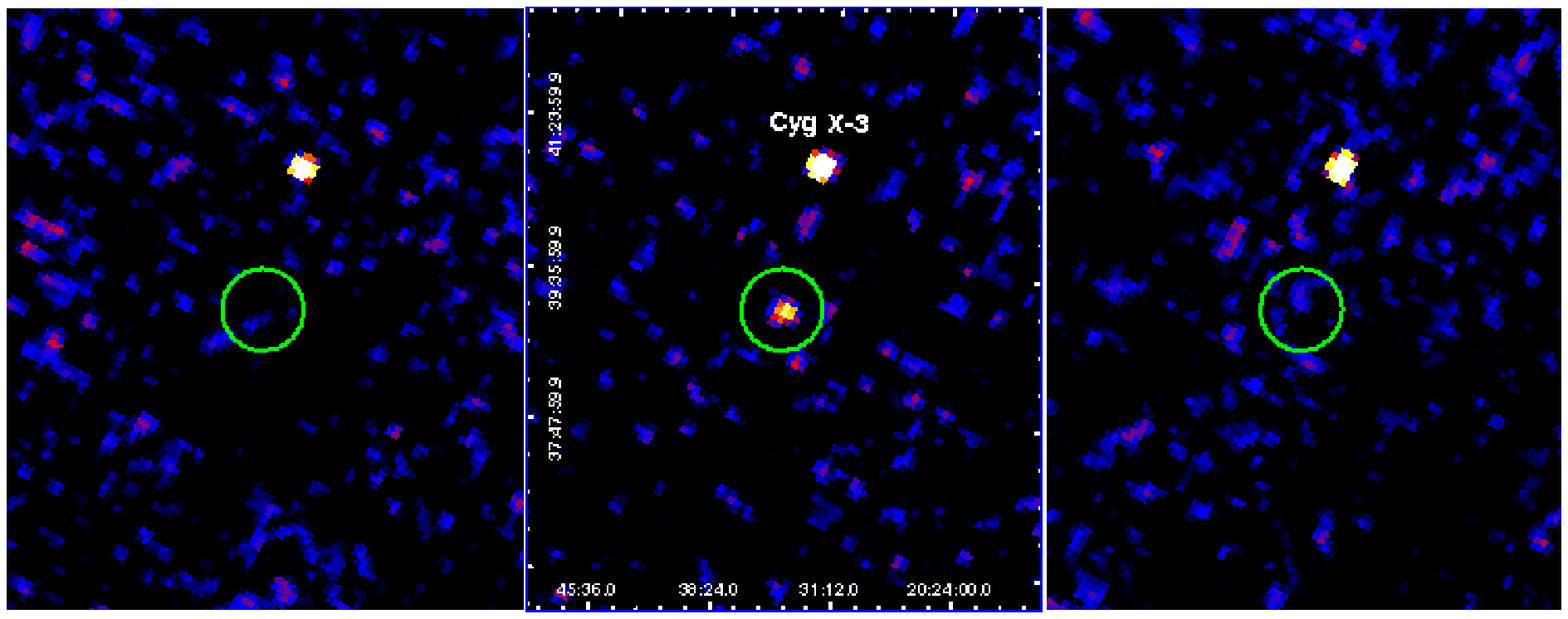,height=4cm,width=9cm}
 \caption{IBIS/ISGRI  ScW image   sequence   (22--60 keV) from number 25 to 27 (revolution 1614) of   the  newly   discovered   transient source   
 IGR~J20344+3913 (encircled).  It was  detected in the middle ScW with a significance of 6.6$\sigma$. The source Cyg X--3 is also detected 
 in the field of view as a bright persistent source.}
\label{fig6}
\end{center}
\begin{center}
\epsfig{file=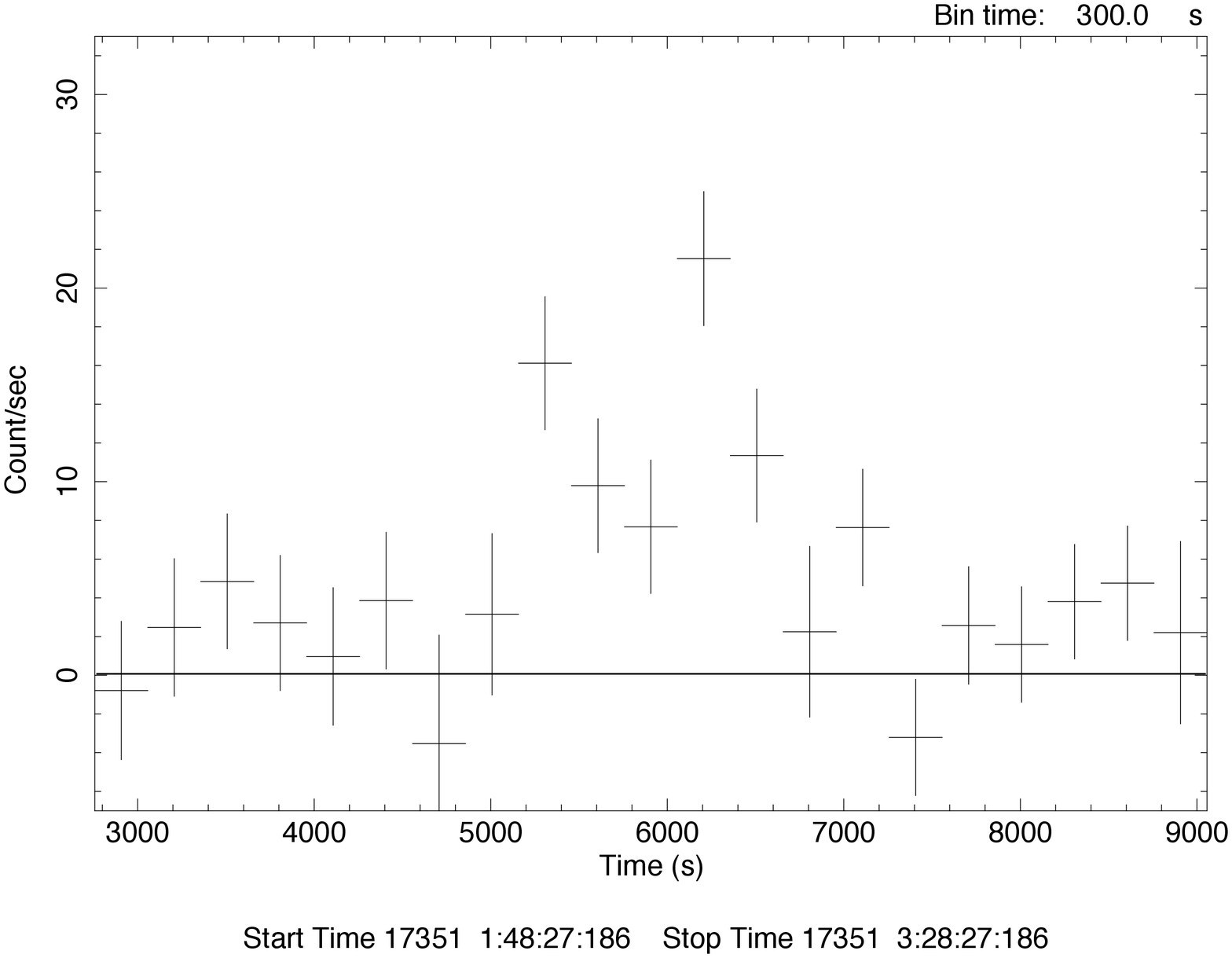,height=6.5cm,width=8.5cm}
 \caption{IBIS/ISGRI 22--60 keV light curve (300 s bin time) of the newly discovered source IGR~J20344+3913. For display purposes aimed at highlighting the transient nature of the source, the light curve was extracted from the three consecutive ScWs number 25 to  27 (revolution 1614) although a source detection was obtained only in ScW number 26.}
\epsfig{file=fig6.ps,height=8.5cm,width=5.5cm,angle=-90}
 \caption{IBIS/ISGRI spectrum  of  IGR~J20344+3913 extracted from Scw number 26 and fitted by a simple power law. The lower panel shows the residuals from the fit.}
\label{fig6}
\end{center}
\end{figure}

\begin{figure}
\begin{center}
\epsfig{file=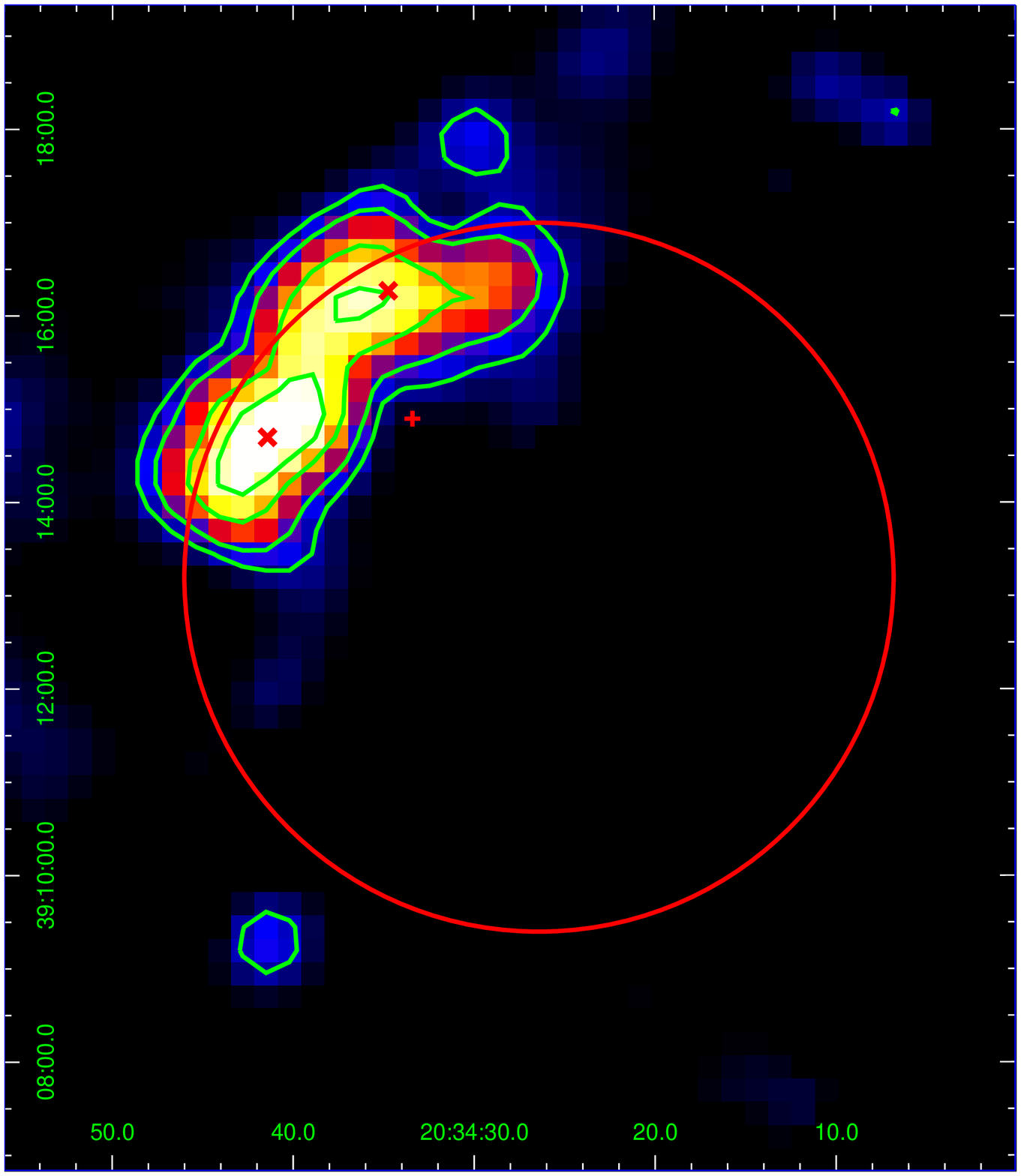,height=8cm,width=7cm}
\caption{NVSS radio map at 20 cm (as taken from the NVSS team web page http://www.cv.nrao.edu/nvss/)
of the sky region containing the newly discovered INTEGRAL source IGR~J20344+3913 (large red error circle). The x and the cross points mark the position of the NVSS  and Green Bank radio sources, respectively.}
\label{fig6}
\end{center}
\end{figure}

We report the discovery of a new fast X-ray transient IGR~J03346+4414
by analyzing IBIS/ISGRI data of the Galactic anticenter region in revolution number 1261.  The source was detected at a significance of 7.2$\sigma$ (22--60 keV) in a single ScW (n. 10) starting at  9 Feb 2013  13:04 (UTC) and ending on the same day at 13:37 (UTC). 

As can be seen from the extracted 22--60 keV light curve (Fig. 1), the duration of the transient activity is only $\sim$ 15 min, it  being characterized by a fast rise ($\sim$ 200 s) followed by a slower decay. At the peak, the source reached a flux of 331$\pm$34 mCrab ($\sim$3.6$\times$10$^{-9}$ erg $^{-2}$ s$^{-1}$).  We point out that the duration of the transient activity ($\sim$ 15 min) is  significantly shorter than that  of the entire ScW  number 10 which contain the detection of the source ($\sim$ 30 min). Bearing this in mind, we performed an imaging analysis with the GTI   by considering only the time interval of the transient activity. By doing so, the source was detected at  $\sim$ 12$\sigma$ level (see Fig. 2) in the energy band 22--60 keV, while no detection was obtained at  higher energies 60--100 keV.  We note that the source was also significantly detected in both the softer (22--30 keV, 6.6$\sigma$) and harder  (30--60 keV, 9.7$\sigma$) energy bands. 

The best position for the source is  RA=53$^\circ$.65 (l=151$^\circ$.25) and Dec=44$^\circ$.24 (b=--9$^\circ$.47) with a 90$\%$ confidence error circle radius equal to 2.1 arcminutes.  No previously known X-ray source is located inside this  error circle according to all the available X-ray catalogs in the HEASARC database. 
Following the source naming convention, we named this newly discovered IBIS/ISGRI source as IGR~J03346+4414.

The IBIS/ISGRI spectrum extracted with the GTI (22--100 keV) is well fitted by a simple power law with $\Gamma$=1.99$\pm0.35$ ($\chi^{2}_{\nu}$=1.05, 9 d.o.f.) and 18--60 keV (20--40 keV)  average flux of 1.9$\times$10$^{-9}$ erg cm$^{-2}$ s$^{-1}$ (1.1$\times$10$^{-9}$ erg cm$^{-2}$ s$^{-1}$). Fig. 3 shows the data-to-model fit with the corresponding residuals. Alternatively, we have also used a thermal model such as  Bremsstrahlung which  provides a good  fit ($\chi^{2}_{\nu}$=0.9, 9 d.o.f.), however the temperature value is  not well constrained (kT=70$^{+66}_{-26}$ keV). 
  
IGR~J03346+4414 is not listed in the  latest published IBIS/ISGRI catalog  (Bird et al. 2016) despite extensive INTEGRAL coverage of 
its sky region  ($\sim$ 1 Ms up to revolution number 1000 considered in Bird et al. 2016) and  this information can be used to infer an upper limit on its persistent emission. By additionally considering the source exposure from  our present dataset ($\sim$ 2 Ms), we can infer a 2$\sigma$ upper limit of $\sim$ 0.2 mCrab or 1.5$\times$10$^{-12}$ erg cm$^{-2}$ s$^{-1}$ (20--40 keV) for persistent emission. When assuming the  source 
peak flux in the same energy  band as measured by IBIS/ISGRI from the outburst reported here,  we can infer a dynamic range greater than at least $\sim$ 2400.

\subsubsection{IGR J20344+3913}

We report the discovery of a new fast X-ray transient found by analyzing public IBIS/ISGRI data of the Cygnus region in 
revolution number 1614.  It was detected at about 6.6$\sigma$ level (22--60 keV) in a single ScW (n. 26) starting at  
25 Nov 2015  02:19 (UTC) and  ending on the same day at 02:52 (UTC). No detection is obtained at  higher energies 60--100 keV. 
We note that the source was barely detected in the softer energy band 22--30 keV (4.3$\sigma$) while the significance 
 was higher  in the harder band 30--60 keV (5.5$\sigma$). Fig. 4 shows the IBIS/ISGRI ScW significance  image sequence (22--60 keV) from number 25 to 27.  
 
Fig. 5 displays  the  22--60 keV light curve which illustrates  the  fast flaring nature of the 
source. The duration of its total activity  was about 0.5 hour,  the source  flared up  reaching a  peak  flux of  138$\pm$19 mCrab ($\sim$1.5$\times$10$^{-9}$ erg cm$^{-2}$ s$^{-1}$).

 The extracted IBIS/ISGRI spectrum (22--100 keV)  is well fitted by a simple power law with 
 $\Gamma$=2.85$^{+0.77}_{-0.65}$ ($\chi^{2}_{\nu}$=1.03, 9 d.o.f.) and 18--60 keV (20--40 keV) 
average flux of 8.1$\times$10$^{-10}$ erg $^{-2}$ s$^{-1}$ (5.1$\times$10$^{-10}$ erg $^{-2}$ s$^{-1}$). Fig. 6 shows the data-to-model fit with the corresponding residuals. A thermal Bremsstrahlung model provides a good fit as well, with $\chi^{2}_{\nu}$=0.9 (9 d.o.f.) and 
kT=28.6$^{+24.7}_{-11.2}$ keV.

The best source position is RA=308$^\circ$.61 (l=78$^\circ$.67) and Dec=39$^\circ$.22 (b=--0.$^\circ$.64) with a 90$\%$ confidence circle radius equal to 3.8 arcminutes. No previously known X-ray source is located inside this error circle according to all the available X-ray catalogs in the HEASARC database. Following the source naming convention, we named this newly discovered IBIS/ISGRI source as IGR~J20344+3913.  

In addition we note that, according to all the available radio catalogs in the HEASARC database, the  bright radio source GB6~J2034+3914 is located inside the IBIS error circle at a distance of 2$'$.1. It is listed in the Green Bank radio catalog (Gregory et al. 1996) with a 6 cm flux of 417$\pm$39 mJy. Further away,  two other bright radio sources (NVSS J203441+391441 and NVSS J203434+391617) listed in the NVSS radio catalog at 20 cm (Condon et al. 1998) are located inside the IBIS error circle at distance (flux)  of  3$'$.3  (84 mJy) and 3$'$.4  (70 mJy), respectively. Fig. 7 shows the IBIS/ISGRI error circle superimposed on the NVSS 20 cm radio map of the sky region.  The x and the cross points mark the position of the NVSS  and Green Bank radio sources, respectively. We note that the two NVSS radio sources are close ($\sim$ 2$'$   from each other) and represent the brightest peak of an extended radio emission as detected at 20 cm. 
The  position of the Green Bank radio source (cross point) is  equally distant ($\sim$ 1$'$.4) from the two NVSS radio sources.  We point out that the 6 cm Green Bank radio survey had an angular resolution of about 3$'$.5 and it detected sources with a positional uncertainty as high as 1$'$. Bearing this in  mind, it cannot be excluded that during its survey the Green Bank telescope detected a radio source (GB6~J2034+3914) which is actually the same source as one or both the NVSS radio objects (NVSS J203441+391441 and NVSS J203434+391617). This possibility is further corroborated by the fact that  GB6~J2034+3914 is listed  in the Green Bank catalog  as   flagged,  to  indicate that there is either significant extension to the source or that the source is partially resolved blend of two or more sources. 
 
IGR~J20344+3913 is not listed in the  latest published IBIS/ISGRI catalog  (Bird et al. 2016)  despite extensive INTEGRAL coverage of 
its sky region  ($\sim$ 3 Ms up to revolution 1000 considered in Bird et al. 2016), this information can be used to infer an upper limit on its persistent emission. By additionally considering the 
source exposure from  our present dataset ($\sim$ 4 Ms), we can infer a 2$\sigma$ upper limit of $\sim$ 0.15 mCrab or 
1.1$\times$10$^{-12}$ erg cm$^{-2}$ s$^{-1}$ (20--40 keV) for persistent emission. When assuming the  source 
peak flux in the same energy  band as measured by IBIS/ISGRI from the outburst reported here,  we can infer a dynamic range greater than at least $\sim$ 1360.

 \begin{table*}
\begin{center}
\caption[]{Swift/XRT (PC) observation log of the two ToOs targeted on the IBIS/ISGRI sky positions.
}
\begin{tabular}{ccccc}
 \hline
\hline
\noalign {\smallskip}
Target   &       Obs. ID          &        Start  Time               &  End Time   &    XRT/PC Exposure Time (s)    \\ 
\hline
\noalign {\smallskip}
IGR~J03346+4414  &        00034387001         &      2016-03-06 00:36:28       &    2016-03-07 00:52:53    &  2932     \\
IGR~J20344+3913  &        00034386001         &      2016-03-04 01:51:47       &    2016-03-04 03:43:54    &  2208     \\
\noalign {\smallskip}
\hline
\label{tab:swlog}
\end{tabular}
\end{center}
\end{table*}

\begin{table*}
\begin{center}
\caption[]{Swift/XRT results of the two ToOs centered on the two IBIS/ISGRI sources, the last row reports the properties of the X-ray source detected in the IGR~J20344+3913 field: 3$\sigma$ upper limits  to the Swift/XRT count rate (0.3--10 keV) of the two IBIS/ISGRI sources are listed in column 2, calculated for a non-detection within the 90\% IBIS error circle.  Assuming the average Galactic column density towards the targets (column 3, Dickey \& Lockman 1990), 
we list in column 4 and 6 the 3$\sigma$ upper limits to the fluxes corrected for the absorption, assuming 
a power law spectrum with a photon index $\Gamma$=1 (UF$_{1}$) and  $\Gamma$=2 (UF$_{2}$), respectively.
L$_{1}$ and L$_{2}$ are the correspondent luminosities, at 10 kpc (IGR~J03346+4414) and 5 kpc (IGR~J20344+3913 and the new XRT source).

}
\begin{tabular}{lcccccc}
 \hline
\hline
\noalign {\smallskip}
Src   &           XRT Rate  (0.3-10 keV)            &           N$_{\rm H}$        &    UF$_{1}$    (0.3-10 keV)  &    L$_{1}$       &    UF$_{2}$     (0.3-10 keV) &    L$_{2}$         \\ 
      &            (counts s$^{-1}$)                &          (cm$^{-2}$)         &    (erg cm$^{-2}$ s$^{-1}$)  &  (erg s$^{-1}$)  &    (erg cm$^{-2}$ s$^{-1}$)  &     (erg s$^{-1}$)        \\
\hline
\noalign {\smallskip}
IGR~J03346+4414 &          $<$ 1.3$\times10^{-2}$           &        2.7$\times10^{21}$    &    $<$ 1.1$\times10^{-12}$ &    $<$    1.3$\times10^{34}$    &    $<$ 8.4$\times10^{-13}$     &    $<$    1.0$\times10^{34}$      \\
IGR~J20344+3913 &          $<$ 1.7$\times10^{-2}$           &        1.2$\times10^{22}$    &    $<$ 2.1$\times10^{-12}$ &    $<$    6.2$\times10^{33}$    &    $<$ 2.2$\times10^{-12}$     &    $<$    6.5$\times10^{33}$      \\
new XRT source &    (8.8$\pm{2.2})\times10^{-3}$          &        1.2$\times10^{22}$      &    1.07$\times10^{-12}$ &        3.2$\times10^{33}$    &     1.12$\times10^{-12}$     &          3.3$\times10^{33}$      \\
\noalign {\smallskip}
\hline
\label{tab:swres}
\end{tabular}
\end{center}
\end{table*}

\section{Soft X-ray observations}

We requested ToO observations at the sky positions of the two new IBIS/ISGRI sources with the $Swift$ satellite, to
look for possible soft X-ray counterparts, and eventually refine their celestial coordinates. 
Indeed, to date no soft X-ray observations (except the $ROSAT$ All Sky Survey) have ever covered the IBIS/ISGRI sky positions.
The two ToOs were performed between 2016 March 4 and 7 (see Table 2 for the summary log) 
in photon counting (PC) mode. 
We used {\em HEASOFT} version 6.18 and the most up to date  calibration files to perform the {\em Swift} data reduction and analysis.


\begin{figure*}
\begin{center}
\includegraphics[height=5.cm,angle=0]{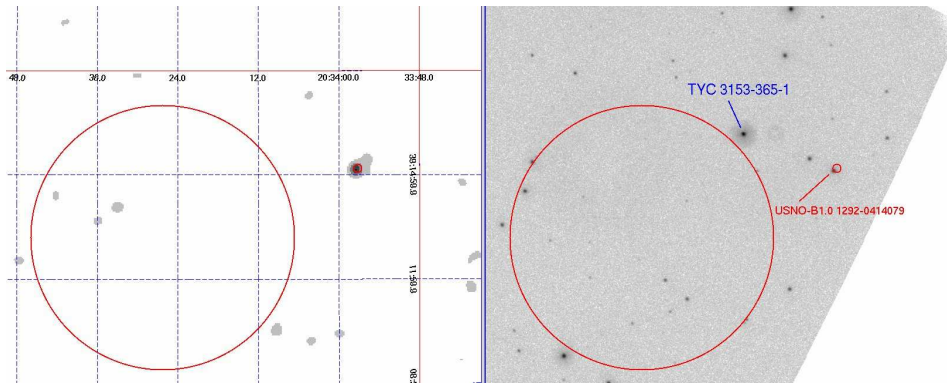}
\end{center}
\caption{$Swift$/XRT observation targeted on the source IGR~J20344+3913. {\em Left panel:} Close-up view of the center of the XRT observation: the large big red circle marks the 3.8$'$ error radius of  IGR~J20344+3913,
while the small red circle (radius, R=7$''$)  indicates the only source detected during this XRT snapshot (the XRT image has been smoothed with a Gaussian filter only for graphical reasons, to better show 
the counts excess). {\em Right panel}: UVOT image, where the same XRT error regions have been displayed in red. The XRT source is positionally coincident 
(within 6$''$) with the star USNO-B1.0~1292-0414079 (2MASS 20335767+3915073, see text). 
Also the bright star TYC 3153-365-1 (VT mag=11.686$\pm{0.115}$) is indicated, to better identify the stellar field. 
}
\label{fig:ima}
\end{figure*}



\subsection{{\itshape Swift} results}


We extracted Swift/XRT sky images in the energy range 0.3-10 keV. 
Using {\em ximage} we searched for soft excesses by means of the tool {\em detect}.
No excesses with a minimum signal-to-noise ratio ({\em snr}) of 3 were found inside the two IBIS/ISGRI error circles.
Therefore, we estimated 3$\sigma$ upper limits to the soft flux, for a source centroid variable within their error circles, 
using the tool {\em sosta} (see Table 3 for these results), which makes use of a local estimate of the background.
Using WebPIMMS and adopting the average absorbing column density along the line of sight (Dickey \& Lockman 1990) we estimated the 3$\sigma$ upper limits (UF) to the soft fluxes corrected for the absorption, 
adopting a power-law model with two values for the photon index ($\Gamma$=1 and $\Gamma$=2).
The results are reported in Table 3, where also the 3$\sigma$ upper limits to the X-ray luminosity are listed.
For IGR~J20344+3913  which is located in the Cygnus region, we assumed a distance of 5 kpc 
in the hypothesis that it resides in 
the Cygnus spiral arm  (Hachisuka et al. 2009,  Kothes et al. 2014). For IGR~J03346+4414,  we assumed a canonical value 
of 10 kpc  for the distance since no assumption can be made on its location arm. 

No other excesses were found in the whole Swift/XRT field-of-view of the two observations, except for the one targeted on IGR~J20344+3913, where a faint source ({\em snr}=3.6) was detected 
at the following sky position: 
R.A. (J2000) = 20:33:57.3, Dec (J2000) = $+$39:15:11.6 (error radius of 7$''$, estimated with the tool {\em xrtcentroid}).
The 0.3-10 keV net source count rate is (8.8$\pm{2.2})\times10^{-3}$~counts~s$^{-1}$ (corrected for the PSF, sampling dead time and vignetting).
In Table~\ref{tab:swres} (last row) we report on the estimated fluxes and luminosities for a power-law continuum. 
This faint X-ray source lies about 6$'$ from IGR~J20344+3913  (well outside its error circle radius of 3.8$'$), 
so it cannot be associated with it. A search in the on-line catalogues at all wavelengths revealed that this new XRT source is positionally coincident (offset of 6$''$)  with the source USNO-B1.0~1292-0414079 (also consistent with 2MASS~20335767+3915073),
showing the following magnitudes: 
J=10.925$\pm{0.025}$,
H=10.445$\pm{0.021}$,
K=10.289$\pm{0.014}$ (2MASS catalogue, Cutri et al. 2003) and 
B1=13.39, R1=12.10, B2=13.90, R2=11.87, I=11.21 (USNO-B1.0 catalogue, Monet et al. 2003). 
In the Guide Star Catalogue (GSC2.3, Lasker et al. 2008) it is classified as a star with a magnitude V=12.47.
Fig.~\ref{fig:ima} shows a close-up view of both XRT and UVOT images, around this faint X-ray source. 
Adopting to its visual magnitude and an X-ray flux of 10$^{-12}$~erg~cm$^{-2}$~s$^{-1}$, we calculated a ratio log(f${_X}$/f${_V}$)=$-1.64$, which 
 is compatible with a stellar origin (Maccacaro et al. 1988), although the X-ray flux is likely overestimated because 
of the assumption of the power-law continuum (Table 3). If the X-ray source is really associated with this star, it
should be located within $\sim$100~pc, to enable a low X--ray luminosity of the order of 1.3$\times10^{30}$ erg s$^{-1}$.


\section{Discussion}
From analysis of archival INTEGRAL data, we have presented IBIS/ISGRI results  on the two newly discovered FXTs
IGR J03346+4414 and IGR J20344+3913. 
 They have been detected only once and never again on February 2013 and November 2015 respectively, being characterized by flaring activity lasting in the range 15--30 minutes and having average flux of the order of  $\sim$ 10$^{-9}$  erg cm$^{-2}$ s$^{-1}$ (22--60 keV).  Recent ToO observations performed with the Swift/XRT satellite failed to detect any quiescent or low level  soft X-ray emission from either  of  the two FXTs, providing stringent 3$\sigma$ upper limits  of the order of a few times 10$^{-12}$  erg cm$^{-2}$ s$^{-1}$ (0.3--10 keV).  Similarly, we obtained stringent upper limits  on their persistent emission above 20 keV as well (of the order of $\sim$ 10$^{-12}$  erg cm$^{-2}$ s$^{-1}$) allowing us to derive high  dynamic range values of 1360 and 2400, respectively. After discovering the two FXTs  analyzing  our IBIS/ISGRI  dataset (revolution number in the range 1535--1629 for the Cygnus region and 918--1262 for the Galactic anticenter), we performed a search at the known source coordinate position  for additional possible flares detected in the  entire IBIS/ISGRI archive  (Paizis et al. 2013) covering revolution number in the range 25--1519 ($\sim$ 12 years of data), i.e. prior to our current dataset. For both FXTs, no detection was found at ScW level  above a significance value of 7$\sigma$ in several different energy bands (17--50 keV,  17--30 keV, 30--50 keV). The  exposure times obtained from such  entire archive revolution 25--1519 were of the order of 
$\sim$ 12 Ms (IGR J20344+3913) and  $\sim$ 5 M s (IGR J03346+4414).

Although we are aware that all the  information reported here is insufficient to firmly identify the nature of the two FXTs, in the following they can be used to obtain some hints and/or indications on the most likely class of X-ray sources to which they  could belong to.

\subsection{IGR J03346+4414}
At first glance,  its location off the Galactic plane (b$\sim$--9.5$^\circ$)  would naturally imply an 
extragalactic blazar nature, however this scenario is strongly weakened by the lack of a  strong radio source inside the error circle according to all the available catalogs in the HEASARC and NED database.  

On the other hand, the fast X-ray transient behavior,  the IBIS/ISGRI spectral shape, the high dynamic range,   all are compatible  with a Galactic  SFXT scenario.  The location of the source off the Galactic plane is apparently irreconcilable  with a HMXB nature since such sources are typically located  on the  Galactic plane in star forming regions. However we note that, although very unlikely, some HMXBs could eventually be located off the plane due to an effect of  perspective if they are particularly nearby ( i.e. distance $<$ 1 kpc) much like the case of the Be HMXB   $\chi$ Per located at b$\sim$--17$^\circ$. Under this assumption,  IGR J03346+4414 would be characterized by a 22--60 keV  average (peak)   outburst luminosity of $<$2$\times$10$^{35}$ erg s$^{-1}$ ($<$4$\times$10$^{35}$ erg s$^{-1}$) while the Swift/XRT  flux upper limit would translate into an average  luminosity  of $<$1$\times$10$^{32}$  erg s$^{-1}$. All such values  are quite low for a SFXT scenario although they could eventually be  still compatible with the weakest hard X-ray flares detected from typical SFXTs  and with the rare quiescent state observed for SFXTs, respectively.    According to the HEASARC database, 40 2MASS sources and 121 USNO-B1.0 optical stars are located within the 2.$'$1 error circle radius of  IGR~J03346+4414.  The brightest near infrared (NIR) source is  2MASS~03344236+4414207 (J=11.716 mag, H=11.126 mag, K=11.005 mag). If we assume a B-type supergiant nature for it to reconcile with both the 2MASS magnitudes and observed colors, it should be affected by an extinction, Av, of 5 magnitudes and it should be located at a large distance of $\sim$20 kpc (which is very unlikely). A much nearer main sequence B star (about 5 kpc) affected by a similar extinction could account for both the IR colors and the faint IR magnitudes, as well. This latter possibility would indicate a Be HMXB nature.  Unfortunately the high number of NIR/optical sources within the large ISGRI error circle prevents us to pinpoint the correct counterpart and we cannot constrain the source nature. To this aim,  it is mandatory to reduce the value of the error circle radius to arcsecond size.

An alternative and interesting Galactic scenario  is  that involving nearby  flare stars. In this case flaring activity  originates in main sequence or
pre-main sequence K-M stars, from plasma  magnetically confined in compact structures in the stellar outer atmosphere and heated  at very hot temperatures  of the order of 10$^7$--10$^8$ K (see Maggio 2008 for a review). Stellar flares in the soft X-ray band 0.2--10 keV have been extensively studied by essentially every major X-ray mission.  Measured soft X-ray fluxes can be as high as $\sim$ 10$^{-8}$  erg cm$^{-2}$ s$^{-1}$ with corresponding 
X-ray luminosities as high as $\sim$ 10$^{33}$  erg s$^{-1}$ for very nearby  stars (i.e. distances typically in  a range from a few parsec to a few  hundreds of parsec). On the contrary,  in the hard X-ray domain  (E$>$ 20 keV) only a few firm events have been detected by the PDS onboard 
BeppoSAX (Schmitt \& Favata 1999, Pallavicini et al. 2000), Swift/BAT (Osten et al. 2007, 2010, 2016, Copete et al. 2008)  and INTEGRAL/IBIS (Bird et al. 2016). This is because the hard X-ray emission is much less intense than soft X-ray (e.g. by a factor of $\sim$ 10$^5$) and the available hard X-ray instruments  have  too limited sensitivity, such that only the largest flares (which are intrinsically rare)  can be unambiguously detected. Hard X-ray emission from stellar flares is mainly thermal in origin, being best interpreted by 
single or multi temperature models of an optically thin plasma (e.g. Bremss, APEC, MEKAL in XSPEC terminology). However  non thermal  emission is also expected in the form of an additional power law tail component which is weak  and very difficult to disentangle from the thermal component, in fact to date non thermal hard X-ray emission  from stellar flares has escaped  firm detections. If we consider the Swift mission, to date it has detected $\sim$ 5 stellar flares which were bright enough to trigger BAT (i.e. flux $> $ 3$\times$10$^{-9}$  erg cm$^{-2}$ s$^{-1}$  in the 15--50 keV band). In most cases only high temperature thermal emission was detected with values as high as $\sim$ 3$\times$10$^{8}$ K, however in at least one case   a possible  excess over the thermal emission component was observed and interpreted as most likely non thermal emission (Osten et al. 2007). 
The  typical durations of the events detected by BAT is from several minutes to several  hours,  conversely in the soft X-ray band stellar flares  can last  even much longer (from few hours to few days).  The duration of the flare detected by INTEGRAL from IGR J03346+4414 ($\sim$ 15 minutes) is compatible with that of stellar flares as typically observed in  the hard X-ray band, moreover its location off the Galactic plane could be eventually  explained by its nearby location. If we assume a reasonable distance 
of 100 pc,  the corresponding 22--60 keV peak (average) X-ray luminosity would be  $\sim$4$\times$10$^{33}$ erg s$^{-1}$ ($\sim$2$\times$10$^{33}$ erg s$^{-1}$) which is consistent with expectations for typical stellar flares  detected in the hard X-ray band. As comparison, the only stellar flare detected by INTEGRAL/IBIS to date is that from the flare star GT Mus (172 parsec distance) which reached a 20--40 keV average luminosity  of $\sim$1.2$\times$10$^{33}$ erg s$^{-1}$ (Bird et al. 2016, 2010).  As for the spectral shape, the hard X-ray spectrum of IGR~J03346+4414 is well fitted by a power law model with a soft photon index of $\sim$ 2,  consistent with expectations  of typical stellar flare  detected in the hard X-ray band.  Alternatively,  a  single temperature model (Bremsstrahlung) also provided  a good  fit with 
a particularly high temperature of  kT$\sim$ 70 keV or T$\sim$ 8$\times$10$^{8}$ K, although the nominal value is poorly constrained in the range  $\sim$ 44--135 keV or (5--15)$\times$10$^{8}$ K. In addition here  we note that an equally good fit ($\chi^{2}_{\nu}$=0.9, 9 d.o.f.)  was also  achieved with a  APEC thermal model,  the temperature was not constrained but its nominal value ($\sim$ 65 keV) was consistent with the Bremsstrahlung measurement. Both temperature values  would clearly be representative of an  "superhot"  thermal plasma (Hudson \& Nitta 1996). 
This high plasma temperature  value could  eventually be deemed unlikely and rejected as unphysically high, but we note that temperatures  as high as $\sim$ 10$^{8}$ K have  been  measured  from many previous  large stellar flares.  For example, a temperature of $\sim$3$\times$10$^{8}$ K  has  been measured with Swift/BAT from the source II PEG  during a huge flare (Osten et al. 2007) as well as from the source DG CVn (Osten et al. 2016). In addition, the Chandra Orion ultra deep project (Getman et al. 2008) allowed the detection of  several  stellar flares with peak plasma  temperatures in excess of $\sim$2$\times$10$^{8}$ K, with the most extreme case even reaching a value of  $\sim$7$\times$10$^{8}$ K. In our specific case, the high measured temperature of $\sim$8$\times$10$^{8}$ K  could be explained either by the hard tail of a "superhot" thermal plasma (among the  hottest thermal emission from possibly a flaring star, to our knowledge), or by a non-thermal component in addition to a thermal plasma emission which we cannot constrain, given the high energy spectrum available (E$>$20 keV). 
 
 \subsection{IGR J20344+3913} 
The fast X-ray transient behavior, IBIS/ISGRI spectral shape (e.g. Sguera et al. 2008), location on the Galactic plane and  high dynamic range strongly suggest that IGR~J20344+3913 is a Galactic SFXT. The 0.3--10 keV   luminosity upper limit ($\sim$6$\times$10$^{33}$ erg s$^{-1}$  at 5 kpc distance) 
is compatible with the source being a SFXT in the  classical  intermediate X-ray state or  during the rarer quiescence.  The inferred duty cycle  is  remarkably low  ($\sim$ 0.01\%) since strong flaring activity  has only been detected by IBIS/ISGRI once  despite its sky region having been extensively observed with a total exposure of  $\sim$ 16 Ms to date (i.e. obtained by summing 12 Ms from entire archive revolution 25--1519 and 4 Ms from our current dataset revolution 1535--1629). Considering that classical SFXTs are characterized by a duty cycle value typically in the range (0.1--5)$\%$ when observed above 20 keV by INTEGRAL (Paizis \& Sidoli 2014), the value  of IGR J20344+3913  is significantly lower than  this range. On  one side this could cast some doubts on its SFXT nature, on the other side it could be due to a marked
scarcity of  detected outbursts which could have been missed by INTEGRAL eventually because of an  observational effect,  i.e.  the source's particularly large distance would allow the detection of only the brightest (and rarest) outbursts , much like the case of the candidate SFXT IGR J18462$-$0311 possibly located at a distance of 11 kpc (Sguera et al. 2013). If we follow this line of reasoning and assume a large distance for  IGR J20344+3913 in the range 5--10 kpc (this being the value derived from the reasonable assumption that it resides  in  the Cygnus spiral arm), then the implied X-ray luminosity at the peak is in the interval (4--20)$\times$10$^{36}$ erg s$^{-1}$  which is effectively representative of the brightest (and rarest) outbursts typically detected from SFXTs. This is particularly true for the highest luminosity value of the order of  $\sim$ 10$^{37}$ erg s$^{-1}$, which tends to favour a large distance of the order of  10 kpc.  A huge number of NIR/optical sources fall within the 3.$8'$ error circle radius of IGR~J20344+3913: 363 2MASS sources and 130 USNO-B1.0 stars.  Among them,  several could be the donor star in a HMXB. For instance, 2MASS 20342790+3913165 is one of the brightest NIR sources (J=10.583 mag, H=9.595, K=9.262 mag)  and it is located at 0.$3'$  from the IGR~J20344+3913  best position.  Its properties would be consistent with a B-type supergiant star located at a distance of 7.6 kpc suffering an extinction of Av=8.6 mag. Conversely, if we assume  that it is a main sequence B star, then it should be located at about 1.9 kpc (with Av=6.5 mag). In conclusion, a SFXT is a viable scenario but unfortunately  the large ISGRI error circle prevents us to pinpoint the correct  NIR/optical counterpart and so we are unable to  constrain the source nature. To this aim it is mandatory to reduce the value of the error circle radius to arcsecond size. Additionally, we note the intriguing presence of a strong radio source (listed  in both NVSS and Green Bank catalogs) inside the IBIS/ISGRI error circle of this candidate SFXT.  However, the possibility that such association could be simply spurious must be taken into account, and to this aim we calculated the  probability  of  finding  a  NVSS radio source  inside the IBIS/ISGRI  error circle by chance. Given the spatial density of NVSS radio sources (as taken from Condon et al. 1998),  we estimated a probability of $\sim$ 0.5 chance coincidences. Such  value is not low enough to strongly support a  real physical  association,  conversely it is likely that the radio 
source is unrelated to IGR J20344+3913.  

As extensively discussed in the previous case of IGR J03346+4414, the interesting scenario of a nearby flare star  is viable for IGR J20344+3913 as well.  Its spectral characteristics (soft power law with $\Gamma$ $\sim$ 2.9 and Bremsstrahlung with  temperature  kT$\sim$ 28 keV), 
plasma temperature (T $\sim$ 3$\times$10$^{8}$ K),  flux/luminosity value, duration of the flaring activity,  are all compatible with expectations  from this interpretation.

An alternative Galactic scenario is that involving Symbiotic X-ray binaries (SyXBs). They are a subclass of low mass X-ray binaries  in
which material is accreted onto a neutron star compact object  from the wind of a late type M giant companion. Only a few SyXBs are known to date (see list in Masetti et al. 2007, Kuranov \& Postnov 2015), they are mainly  characterized by relatively low X-ray luminosity both at soft and hard 
X-rays (typically L$_x$$\sim$10$^{32}$--10$^{35}$ erg s$^{-1}$)  and show  long as well as short term X-ray variability. Occasionally, they can  exhibit flaring activity typical of neutron stars accreting matter from a stellar wind. The majority of known SyXBs have been detected by INTEGRAL as  persistent hard X-ray sources,  as such they are listed in several  IBIS catalogs published to date (e.g. Bird et al. 2007, 2010, 2016) with typical  fluxes of the order of $\sim$10$^{-10}$--10$^{-11}$ erg cm$^{-2}$ s$^{-1}$ (20--40 keV). This is  at odds with the  INTEGRAL/IBIS 
non detection of persistent emission from IGR~J20344+3913 (20--40 keV upper  limit of 1.1$\times$10$^{-12}$ erg cm$^{-2}$ s$^{-1}$) 
despite its  sky region has been extensively exposed (i.e. the source  is not listed in the latest published INTEGRAL/IBIS catalog Bird et al. 2016). 
In addition, the extremely short and energetic flare observed from IGR J20344+3913 is not consistent with expectations for typical  duration  observed for SyXBs in the hard X-ray band  (i.e. from several days on).  

Alternatively IGR~J20344+3913  could be a blazar  behind the Galactic plane,  hypothesis supported by the  presence of a bright  radio source inside its error circle (420 mJy at 6 cm) as  listed in the Green Bank radio catalog.  The Swift/XRT  flux upper limit (2.2$\times$10$^{-12}$  erg cm$^{-2}$ s$^{-1}$)  is compatible with this blazar interpretation since it is notably in the range of soft X-ray flux values (0.6--8)$\times$10$^{-12}$  erg cm$^{-2}$ s$^{-1}$  typically measured from blazars detected by INTEGRAL (Malizia et al. 2016), and  therefore it is likely that even small  decrements of the X-ray flux (which is likely to happen given the variable nature of blazar) could have been sufficient to hamper the detection by Swift/XRT in a very short exposure.   However there are several shortcomings for such blazar scenario: i) as reported above, it is likely that the association between 
IGR~J20344+3913 and the radio source is spurious;  ii)  the rather short duration of the IBIS/ISGRI flare  ($\sim$ 30 minutes)  would be  among the shortest  flaring episodes ever detected at hard X-rays from a blazar. In the literature,  only one  similar case is reported, i.e. a $\sim$ 30 minutes flare  from a firm blazar likely detected by INTEGRAL (Foschini et al 2006);  iii) our reported  dynamic range of IGR~J20344+3913 at hard X-rays (about three orders of magnitude) is extreme if compared to blazars, in fact as from observations of blazar X-ray variability their typical dynamic range  is fairly small (from a few to a very few tens in the X-ray band). Higher blazar dynamic range values (up to about two orders of magnitude)  have been achieved 
at gamma-ray energies only in very exceptional cases, like the remarkable blazar 3C~454.3 which is to date the most variable and bright gamma-ray blazar detected by both Fermi and AGILE (Vercellone 2012).

\section{Conclusions}

We reported on the   discovery of two new Fast X-ray Transients (FXTs) from analysis of  archival INTEGRAL data. The main spectral and temporal characteristics of  both IGR~J03346+4414  and IGR~J20344+3913 are  best compatible with a Galactic origin such as  a SFXT or a nearby flare star. Conversely,  the extragalactic  blazar scenario presents some shortcomings. Regardless of their nature,  their peculiar characteristics (e.g. high plasma temperature for IGR~J03346+4414 ) as well as  their unusually short  and bright  outbursts make them particularly interesting. 
This kind of source is very difficult  to discover and characterize due to the very transitory nature and especially the very low duty cycle. The instrumental characteristics of IBIS/ISGRI
onboard INTEGRAL  are particularly suited in  serendipitously detecting and discovering such short duration random events. 
It seems  plausible that  other such sources wait to be discovered, further exploitations of the entire INTEGRAL data archive may yield additional discoveries  of this kind of interesting X-ray transients.

\section*{Acknowledgments}
We thank the anonymous referee for useful comments which helped us to improve the quality of this paper.
We thank the {\em Swift} team, the PI, the duty scientists and science planners
 for making the two ToO observations reported here possible.  
We acknowledge financial support from the Italian Space Agency via INTEGRAL
ASI/INAF agreement n. 2013-025.R.0, and the grant from PRIN-INAF 2014, 
``Towards a unified picture of accretion in High Mass X-Ray Binaries''.
This work has made use of the INTEGRAL archive developed at INAF-IASF Milano (http://www.iasf-milano.inaf.it/\textasciitilde{}ada/GOLIA.html). 
This  research has made use of data and/or software provided by the 
High Energy Astrophysics Science Archive Research Center (HEASARC), which is a 
service of the Astrophysics Science Division at NASA/GSFC and the 
High Energy Astrophysics Division of the Smithsonian Astrophysical Observatory.

{}

\end{document}